\documentclass[3p,a4paper,twocolumn]{elsarticle}

\usepackage{amsfonts}
\usepackage{amsmath}
\usepackage{amssymb}
\usepackage{graphicx}
\begin{document}

\begin{frontmatter}

\title{Long-range quantum discord in critical spin systems}

\author{J. Maziero\fnref{fn2}}
\ead{jonasmaziero@gmail.com}
\fntext[fn2]{Present address: Departamento de F\'isica, Universidade Federal de Santa Maria, CEP 97105-900, Santa Maria, RS, Brazil}

\author{L. C. C\'{e}leri}
\ead{lucas.celeri@ufabc.edu.br}

\author{R. M. Serra}
\ead{serra@ufabc.edu.br}

\address{Centro de Ci\^{e}ncias Naturais e Humanas, Universidade Federal do ABC, R.
Santa Ad\'{e}lia 166, Santo Andr\'{e}, 09210-170, S\~{a}o Paulo, Brazil}

\author{M. S. Sarandy\corref{cor1}\fnref{fn1}}
\ead{msarandy@if.uff.br}
\address{Instituto de F\'{\i}sica, Universidade Federal Fluminense,
Av. Gal. Milton Tavares de Souza s/n, Gragoat\'a, 24210-346, Niter\'oi, RJ, Brazil.}
\cortext[cor1]{Corresponding author}
\fntext[fn1]{Tel.: +55-21-2629-5802 / Fax: +55-21-2629-5887 }

%-------------------------------------------------------------------------------------------------------------------------------

\begin{abstract}
We show that quantum correlations as quantified by quantum discord can characterize quantum phase transitions by exhibiting 
nontrivial long-range decay as a function of distance in spin systems. This is rather different from the behavior of pairwise 
entanglement, which is typically short-ranged even in critical systems. In particular, we find a clear change in the decay rate 
of quantum discord as the system crosses a quantum critical point. We illustrate this phenomenon for first-order, second-order, 
and infinite-order quantum phase transitions, indicating that pairwise quantum discord is an appealing quantum correlation function 
for condensed matter systems.

\end{abstract}

\begin{keyword}
Quantum discord \sep Entanglement \sep Quantum phase transitions \sep Critical spin systems
\end{keyword}

\end{frontmatter}

%---------------------------------------------------------------------------------------------------------------------------------

\section{Introduction} 

A quantum phase transition (QPT) \cite{Sachdev} is primarily characterized by a qualitative sudden change in the ground state of 
an extended quantum system as an external parameter or an internal coupling is continuously varied. QPTs occur effectively at zero temperature and its typical quantum fluctuations are believed to be due to genuine quantum correlations. In recent years, concepts from quantum information theory have been extensively applied to the study of QPTs in quantum many-body systems \cite{Amico1,Amico:09}. In particular, the existence of quantum correlations has usually been inferred by the presence of entanglement among parts of a system. Indeed, entanglement displays a rather interesting behavior at QPTs, being able to indicate a quantum critical point (QCP) through nonanalyticities inherited from the ground state energy \cite{Osterloh1,Wu:04}. This behavior is already observed from pairwise measures of entanglement as given, e.g., by concurrence \cite{Wootters:98} and negativity \cite{Vidal:02a}. 

Although pairwise entanglement measures usually exhibit scaling behavior at a QCP, they are typically exponentially short-ranged \cite{Osterloh1}. To some extent, long-distance pairwise entanglement may be engineered in several many-body systems by conveniently setting microscopic parameters. However, this adjustment does not coincide with QCPs in those systems nor critical scaling of entanglement with distance is observed \cite{Venuti:06}. This is somewhat surprising since one expects a slow decay of quantum correlations in a quantum critical region. In this Letter, we investigate this problem by quantifying quantum correlations through quantum discord (QD) \cite{Ollivier1,review}. In particular, we show that QD provides the expected long-range behavior of quantum correlations for several spin chains exhibiting QPTs. Such a behavior is remarkable since, even though QD has been shown to be non-vanishing for almost all quantum states \cite{Ferraro:10}, its decay pattern as a fu
 nction of distance is unresolved in general grounds. We illustrate our results for the transverse field XY chain, where long-distance QD is achieved due to the onset of magnetic order, as well as for the XXZ chain in the presence of domain walls, where the long-range behavior of quantum correlations is obtained as a consequence of the polynomial decay of QD.

%--------------------------------------------------------------------------------------------------------------------------------

\section{Quantum discord} 

Before introducing quantum discord, let us call up some concepts from classical information theory (CIT). In CIT the uncertainty about a random variable $A$, which can assume the values $a$ (from a set $\mathcal{A}$) with corresponding probability $p_{a}:= \mathrm{Pr}(A=a)$, is given by the Shannon's entropy $H(A)=-\sum_{a}p_{a}\log_{2}p_{a}$. The uncertainty about two random variables $A$ and $B$ taken together reads $H(AB)=-\sum_{a,b}p_{a,b}\log_{2}p_{a,b}$, with $\{p_{a,b}:=\mathrm{Pr}(A=a,B=b)\}$ being the joint probability distribution. The total amount of correlation between $A$ and $B$ is given by the difference in the uncertainty about $A$ before and after $B$ is known, \textit{i.e.}, $\mathcal{J}(A:B)=H(A)-H(A|B)$, where $H(A|B)=-\sum_{a,b}p_{a,b}\log_{2}p_{a|b}$ is the conditional entropy, with $p_{a|b}$ standing for the probability for $A=a$ given that $B=b$. From Bayes' rule, $p_{a|b}=p_{a,b}/p_{b}$, we can rewrite $\mathcal{J}(A:B)$ in the equivalent
form $\mathcal{I}(A:B)=H(A)+H(B)-H(AB)$.

In the quantum domain, the uncertainty over the state $\rho$ is given by the von Neumann's entropy 
$S(\rho)=-\mathrm{tr}(\rho\log_{2}\rho)$. Thus, a straightforward generalization of $\mathcal{I}(A:B)$ to the quantum realm readily follows as
\begin{equation}
I(\rho_{ab}):=S(\rho_{a})+S(\rho_{b})-S(\rho_{ab}).\label{IQ2}
\end{equation}
The quantity $I(\rho_{ab})$ is called quantum mutual information and is a well established information-theoretic measure of the total (quantum plus classical) correlations in a quantum state \cite{QMI}. On the other hand, the average uncertainty about the state of the system $A$ after a complete set of projective measurements $\{\Pi_{j}\}$ is performed on system B is given by $S_{\Pi}(\rho_{a|b})=\sum_{j}p_{j}S(\rho_{a}^{j})$, where
$p_{j}=\mathrm{tr}_{b}(\Pi_{j}\rho_{b})$ and $\rho_{a}^{j}=\mathrm{tr}_{b}(\Pi_{j}\rho_{ab}\Pi_{j})$. Thus, a quantum version of $\mathcal{J}(A:B)$, independent of measurement \textquotedblleft direction\textquotedblright\ in the Hilbert's space, can be defined as
\begin{equation}
J(\rho_{ab}):=S(\rho_{a})-\min_{\{\Pi_{j}\}}S_{\Pi}(\rho_{a|b}).\label{IQ1}
\end{equation}
In Ref. \cite{Ollivier1}, Ollivier and Zurek noticed that, while the classical expressions for mutual information given by $\mathcal{J}(A:B)$ and $\mathcal{I}(A:B)$ are equivalent, their quantum generalizations (\ref{IQ2}) and (\ref{IQ1}) can be different 
depending on the state of the system. This difference originated the quantum discord $Q(\rho_{ab})$, which reads
\begin{equation}
Q(\rho_{ab}):=I(\rho_{ab})-J(\rho_{ab}).\label{QD}
\end{equation}

The QD is a non-negative asymmetric quantity that vanishes only for states encoding a joint classical probability distribution,
\begin{equation}
\sum_{i,j}p_{ij}|i_{a}\rangle\langle i_{a}|\otimes|j_{b}\rangle\langle j_{b}|,  
\end{equation}
where $\{|i_{a}\rangle\}$ ($\{|j_{b}\rangle\}$) form an orthonormal basis for the system $A$ ($B$) (see Refs. \cite{Maziero2,Xu-pla} for a discussion about the symmetry properties of quantum correlation quantifiers). QD is a measure of correlations based on information-theoretic concepts and is intended to capture all quantum correlations present in a quantum state \cite{Ollivier1}. Recently, it has received a great deal of attention, exhibiting remarkable behavior under decoherence \cite{Maziero3-1}-\cite{Yao}, being experimentally tested \cite{QDexp1}-\cite{Auccaise}, and displaying applications in several contexts \cite{QDapp1}-\cite{QDapp8}. Concerning QPTs, QD has been considered as an indicator of QCPs in several systems, having succeeded in this task even in situations where entanglement fails \cite{Maziero1}-\cite{QD-QPTs5}.

%-------------------------------------------------------------------------------------------------------------------------------

\section{The XY model} 

In order to investigate the scaling behavior of QD as a function of distance, let us begin by considering a chain of spin-$1/2$ particles anisotropically interacting in the xy spin plane and subjected to a magnetic field in the z-direction. This system is described by
the XY model, governed by the following normalized Hamiltonian
\begin{eqnarray}
\hspace{-1cm}H&=&-\frac{\lambda}{2}\sum_{i=0}^{N-1} \left[ (1+\gamma)\sigma_{i}^{x}\sigma_{i+1}^{x} +(1-\gamma)\sigma_{i}^{y}\sigma_{i+1}^{y}\right] \nonumber \\
&&-\sum_{i=0}^{N-1}\sigma_{i}^{z}, 
 \label{XY}
\end{eqnarray}
with $N$ being the number of spins in the chain, $\sigma_{i}^{m}$ the $i$-th spin Pauli operator in the direction $m=x,y,z$ and periodic boundary conditions are assumed, i.e., $\sigma_{N}^{m}=\sigma_{0}^{m}$. The anisotropy $\gamma$ is constrained to the range $0\le\gamma\le1$. For $\gamma\rightarrow0$, the model reduces to the XX model whereas for all the interval $0<\gamma\le1$ it belongs to the Ising universality class, reducing to the transverse field Ising model at $\gamma=1$. The dimensionless parameter $\lambda$ is proportional to the reciprocal of the external transverse magnetic field. For $\lambda=1$, a quantum critical line 
takes place for any value of $\gamma$ in the range considered in this work.

The exact analytical solution of the XY model in the thermodynamical limit ($N\rightarrow\infty$) is well known \cite{XYsol1,XYsol2}. The Hamiltonian (\ref{XY}) can be diagonalized via a Jordan-Wigner map followed by a Bogoliubov transform. By considering the thermal state at finite temperature and taking into account the $Z_2$ symmetry of the XY model, namely, the invariance under parity transformation $\bigotimes_{i=1}^{N} \sigma^z_i$, the reduced state for the spins $0$ and $n$ reads \cite{Osborne1}
\begin{eqnarray}
\rho_{0n}=\frac{1}{4}\{\mathbf{I}^{0n}+\langle\sigma^{z}\rangle(\sigma_{0}^{z}+\sigma_{n}^{z}) \nonumber \\ 
+\sum_{i={x,y,z}}\langle\sigma_{0}^{i}\sigma_{n}^{i}\rangle\sigma_{0}^{i}\sigma_{n}^{i}\},
\label{RDO}
\end{eqnarray}
where $\mathbf{I}^{0n}$ is the identity operator acting on the joint state space of the spins $0$ and $n$. Due to the fact that the system is translationally invariant, the reduced state (\ref{RDO}) depends only on the distance $n=|i-j|$ between spins $i$ and $j$. The magnetization density $\langle\sigma^{z}\rangle$ as well as the two-point correlation functions $\langle\sigma_{0}^{i}\sigma_{n}^{i}\rangle$ can be directly obtained from the exact solution of the model \cite{XYsol1,XYsol2} (their expressions are made explicit in Appendix A). We will use throughout this Letter the zero temperature limit ($T\rightarrow0$) of the reduced thermal state given in Eq.~(\ref{RDO}), which is called thermal ground state. This unbroken state was shown to provide an exact description of the critical behavior of entanglement as well as its scaling in finite systems \cite{Osterloh1,Osborne1,SSBe}. 
The total information shared by the spins in the state (\ref{RDO}) is given by 
\begin{equation}
I(\rho_{0n})=S(\rho_{0})+S(\rho_{n})-S(\rho_{0n}), 
\label{IXY}
\end{equation}
with 
\begin{eqnarray}
S(\rho_{0})=S(\rho_{n})= \nonumber \\
-\sum_{i=0}^{1}\frac{1+(-1)^{i}\langle\sigma^{z}\rangle}{2}\log_{2}\frac{1+(-1)^{i}\langle\sigma^{z}\rangle}{2}
\end{eqnarray}
and
\begin{equation}
S(\rho_{0n})=\sum_{i=0}^{1}(\xi_{i}\log_{2}\xi_{i}+\eta_{i}\log_{2}\eta_{i}), 
\end{equation}
with \\
$\xi_{i}=(1+\langle\sigma_{0}^{z}\sigma_{n}^{z}\rangle)/4$\\
$+(-1)^{i}\sqrt{(\langle\sigma_{0}^{x}\sigma_{n}^{x}\rangle-\langle\sigma_{0}^{y}\sigma_{n}^{y}\rangle)^{2} +4\langle\sigma^{z}\rangle^{2}}]/4$,
\\ and \\
$\eta_{i}=\left[1-\langle\sigma_{0}^{z}\sigma_{n}^{z}\rangle+(-1)^{i}(\langle\sigma_{0}^{x}\sigma_{n}^{x}\rangle+\langle\sigma_{0}^{y}\sigma_{n}^{y}\rangle)\right]/4$.

Following \cite{Maziero1}, we numerically verified that the minimum in Eq. (\ref{IQ1}) is attained, for all values of $\lambda$, $\gamma$, and $n$ considered in this Letter, by the following set of projectors:
$\{|+\rangle\langle+|,|-\rangle\langle-|\}$, with $|\pm\rangle=(\left\vert\uparrow\right\rangle \pm\left\vert \downarrow\right\rangle )/\sqrt{2}$, where $\{\left\vert \uparrow\right\rangle ,\left\vert \downarrow\right\rangle \}$
are the eigenstates of $\sigma^{z}$. Thus one obtains
\begin{equation}
J(\rho_{0n})=H_{bin}(p_{1})+H_{bin}(p_{2}),
\end{equation}
where $H_{bin}(x)$ is the binary entropy 
\begin{equation}
H_{bin}(x)=-x\log_{2}x-(1-x)\log_{2}(1-x)
\end{equation}
and
\begin{eqnarray}
p_{1}&=&\frac{1}{2}\left(1+\langle\sigma^{z}\rangle\right), \nonumber \\
p_{2}&=&\frac{1}{2}\left(1+\sqrt{\langle\sigma_{0}^{x}\sigma_{n}^{x}\rangle^{2}+\langle\sigma^{z}\rangle^{2}}\right).
\label{finalXY}
\end{eqnarray} 
This provides, therefore, an analytical expression for evaluating QD for a distant pair described by $\rho_{0n}$. We note that the algorithm proposed in Ref. \cite{Ali} to evaluate QD in X states is indeed valid in our case if 
$|\langle\sigma^{x}_{0}\sigma^{x}_{n}\rangle| \geq |\langle\sigma^{y}_{0}\sigma^{y}_{n}\rangle|$ \cite{Chen}. Using the analytical solution for the XY model, one can verify this inequality for the whole range of values of $\lambda$, $\gamma$, and $n$ considered in this Letter. Moreover, as shown in \cite{Chen}, the difference between the QD obtained using the algorithm of Ref. \cite{Ali}  and that obtained by maximizing over all measurements is very feeble. Thus, by applying Eqs.(\ref{IXY})-(\ref{finalXY}), we can reliably analyze the decay pattern of correlations in the XY chain.

\begin{figure}[!]
\begin{center}
{\includegraphics[angle=0,scale=0.60]{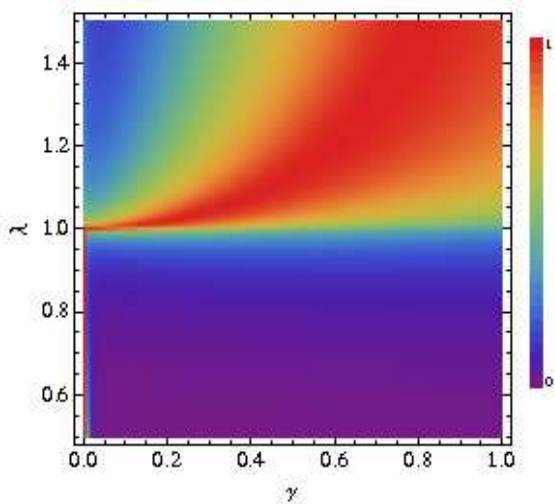}}
\caption{\normalsize{(Color online.) Pattern of decay of quantum discord with the distance between the spins sites as a function of $\lambda$ and $\gamma$. Colors toward the red indicate a slower decay of quantum discord with distance (see the text for details).}}
\label{f1}
\end{center}
\end{figure}

\begin{figure}[!]
\begin{center} 
{\rotatebox{270}{\includegraphics[height=7.8cm,width=19.2cm]{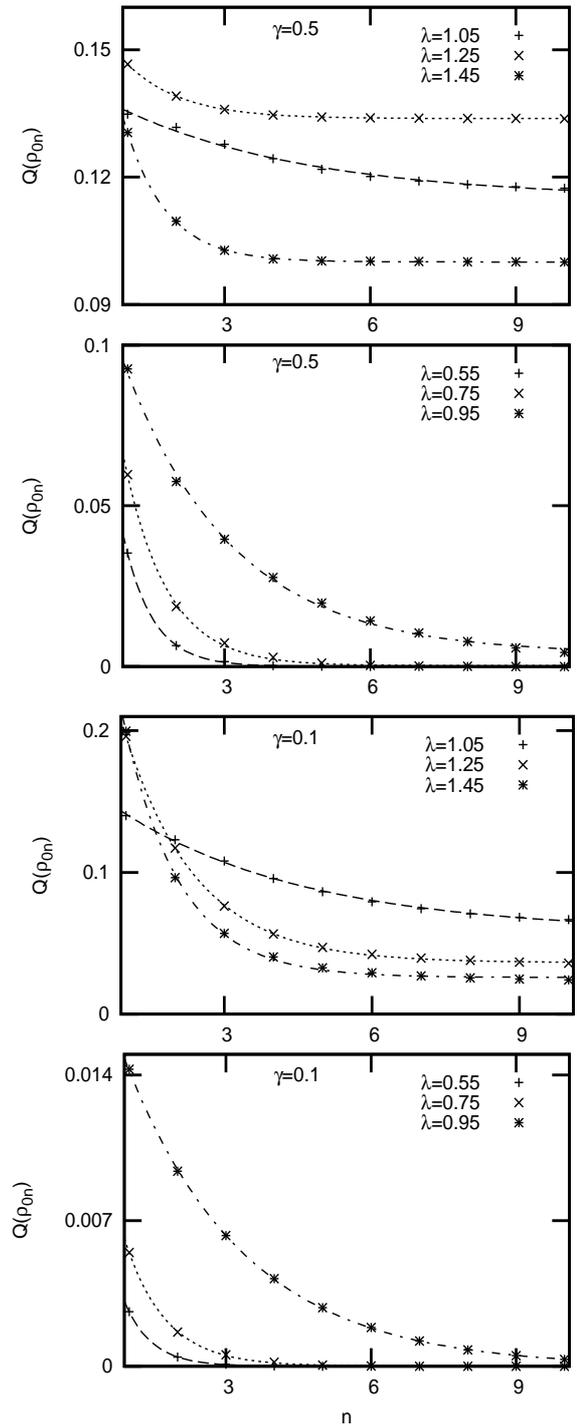}}}
\caption{\normalsize{Decay of quantum discord with the distance between the spins sites for some values of $\lambda$ and $\gamma$. The points are the computed values of QD and the lines are the exponential fits (see the text for details).}}
\label{f2}
\end{center}
\end{figure}

A change in the range of correlations is a typical indication of a QPT in a many-body system. Indeed, such a change is shown in Fig.~\ref{f1}, where we plot a pattern of decay of QD as function of $\lambda$ and $\gamma$. This figure was constructed by computing the ratio between the quantities $\sum_{n=1}^{M}Q(\rho_{0n})$ and $MQ(\rho_{01})$, with $M=10$, for each pair 
($\gamma$,$\lambda$). In other words, we compare the actual area under the curve $Q(\rho_{0n})$ with the bare case where QD remains constant with distance. Although Fig.~\ref{f1} itself does not reveal the decay rate of QD, it clearly shows a qualitative change in the distance behavior of QD as the system passes through the critical line $\lambda = 1$. Indeed, a slower decay of QD is exhibited in the region $\lambda > 1$, where magnetic order takes place. This change in the range of pairwise quantum correlations at 
$\lambda = 1$ is made explicit in Fig.~\ref{f2}, where we plotted QD for a spin pair as a function of the distance $n$. The curves in Fig.~\ref{f2} are the exponential fits of quantum discord. We observe that, for both examples of anisotropies considered, $\gamma=0.1$ and $\gamma=0.5$, the decay of QD with distance can be well fitted by an exponential function $a+b\exp(-cn)$, where $a$, $b$, and $c$ are constants. Nevertheless, we notice that, while for $\lambda<1$ QD vanish exponentially, in the cases where $\lambda>1$, we obtain a constant long-distance value for QD that depends only on $\gamma$ and $\lambda$.

%-------------------------------------------------------------------------------------------------------------------------------

\section{XXZ chain in the presence domain walls} 

In order to consider QPTs of first-order and infinite-order (see, e.g., Ref.~\cite{Rulli}), let us analyze the XXZ spin-1/2 chain in the presence of a boundary magnetic field generating domain walls, whose Hamiltonian reads~\cite{sch,alcaraz}
\begin{eqnarray}
H_{xxz}=-\frac{J}{2}\sum_{i=1}^{N-1}(\sigma_{i}^{x}\sigma_{i+1}^{x}+\sigma_{i}^{y}\sigma_{i+1}^{y}) \nonumber\\
-\frac{\Delta}{2}\sum_{i=1}^{N-1}\sigma_{i}^{z}\sigma_{i+1}^{z}-h(\sigma_{1}^{z}-\sigma_{N}^{z}),
\label{hdw}
\end{eqnarray}
where the coupling $J$ and $\Delta$ are exchange parameters, with the effective magnetic field $h>0$ representing the interactions of the spins with the boundary surfaces. In order to focus the discussion on the values of the anisotropy and the magnetic field, let us assume, without loss of generality, $J=1$. For $\Delta \le -1$, the model is in an antiferromagnetic phase while for $-1 < \Delta < 1$ the model is in a disordered critical (gapless) region. The model exhibits an infinite-order QPT at the antiferromagnetic point $\Delta = -1$ and a first-order QPT at ferromagnetic point $\Delta = 1$. Moreover, for $\Delta\ge 1$, as shown in Ref.~\cite{alcaraz}, the model presents a further first-order QPT governed by the critical field 
\begin{equation}
h_{c}=\frac{1}{2}\sqrt{\Delta ^{2}-1},
\end{equation}
which separates two quantum phases for a chain of arbitrary length: a ferromagnetic ground state 
($h<h_{c}$) and kink-type ground state ($h>h_{c}$). Remarkably, $h_c$ provides the exact location of the (first-order) phase transition for chains of any size, the critical field remaining fixed as  the number of sites is changed. 

\begin{figure}[!]
\begin{center}
{\includegraphics[angle=0,scale=0.31]{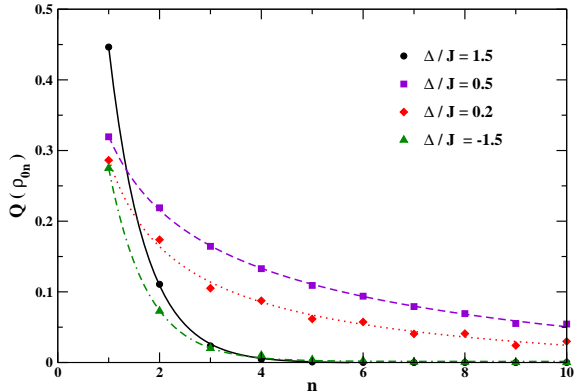}}
\caption{\normalsize{(Color online.) Behavior of QD as a function of the distance for the XXZ chain with domain walls for $N = 22$ sites and magnetic field $h/J=5.0$. The fitting curves in the gapless region $-1 < \Delta/J < 1$  display the form $Q(n) = a + b\, n^{-c}$ (power-law decay) whereas in the antiferromagnetic and kink-type regions the form is $Q(n) = a + b\, \exp(-c\, n)$ (exponential decay), with $a$, $b$, and $c$ constants.}}
\label{f3}
\end{center}
\end{figure}

For the Hamiltonian given by Eq.~(\ref{hdw}), we numerically compute QD for a finite-size chain by exact diagonalization. In the regime of temperature $T=0$, we take the system in its ground state. From the ground state wave-function, we then obtain the two-spin reduced density matrices for arbitrarily distant pairs. Since the system exhibits $Z_2$-symmetry, the numerical evaluation of 
Eq.~(\ref{QD}) can be performed similarly as in the XY model. Results for a chain with $N=22$ sites is shown in Fig.~\ref{f3}, where QD is taken for the two-spin density operator $\rho_{0n}$, with $\rho_{0n}$ standing for the pair ($N/2$, $N/2\,+\,n$). Remarkably, 
a power-law decay is observed in the gapless region $-1 < \Delta < 1$ while exponential decay occurs for the antiferromagnetic and kink-type phases (for the ferromagnetic phase, QD is vanishing~\cite{QD-QPTs1,QD-QPTs2}). This observation keeps unchanged for distinct magnetic fields $h$ as long as $h$ keeps the system in a fixed quantum phase.  

%-------------------------------------------------------------------------------------------------------------------------------

\section{Conclusion} 

In summary, we have investigated the decay of pairwise quantum correlations (as given by quantum discord) as a function of distance along a spin chain. We have found distinct behaviors for the decay of QD as a function of distance as QCPs are crossed in the quantum phase diagram. In particular, the onset of macroscopic order has been shown to be accompanied by the development of long-distance QD in the XY model. For the XXZ chain with domain walls, we have shown that the critical region exhibits power-law decay for QD whereas the gapful (antiferromagnetic and kink-type) phases display exponential decay. With regard to the fact that the set of zero-QD states has zero volume in the state space \cite{Ferraro:10}, we observe that the possible existence of quantum discord for distant sites does not imply \textit{a priori} in a power-law decay or in a decay to a constant value of QD. In fact, this is an interesting and surprising result, which is in contrast with the behavior of pa
 irwise entanglement, which is typically short-ranged (see, e.g., the discussion in Ref. \cite{Osterloh1} and the Figs. 2(e) and 2(f) in Ref. \cite{Maziero1}). This characterization of QPTs in a condensed matter system from the point of view of information theory is an important aspect of the cross-fertilization of these two fields, turning out to be possible only as we consider measures of quantumness that go beyond the entanglement-separability paradigm. In particular, a promising result presented here is the asymptotically constant behavior of QD in the XY model. Indeed, since QD has been recognized as a resource in several contexts, it is potentially relevant to further investigate the possible role of this long-range behavior of QD in quantum communication protocols \cite{Bose} and its possible effects in biological systems \cite{Bio}. Further appealing directions for future research are the investigation of the critical aspects of QD (correlation length, critical expo
 nents, etc.) and the behavior of multipartite measures of quantum cor
relations (see, e.g., Ref. \cite{Modi:10,Rulli:11,Paternostro:11}) at QPTs.

\section*{Acknowledgments}
We are grateful for the funding from UFABC, CNPq, CAPES, FAPESP, FAPERJ, and INCT-IQ.

\appendix
\section{Magnetization density and correlation functions}
 
The transverse magnetization density is given by \cite{XYsol1}
\begin{equation}
\langle\sigma^{z}\rangle=-\int_{0}^{\pi}\frac{\left(  1+\lambda\cos
\phi\right)  \tanh\left(  \beta\omega_{\phi}\right)  }{2\pi\omega_{\phi}}%
d\phi\text{,}%
\end{equation}
where
\begin{equation}
\omega_{\phi}=\frac{1}{2}\sqrt{\left(\gamma\lambda\sin\phi\right)^{2}+\left(1+\lambda\cos\phi\right)^{2}}
\end{equation}
and $\beta=1/k_{B}T$ with $k_{B}$ being the Boltzmann's constant and $T$ the absolute temperature. 

The two-point correlation functions read \cite{XYsol2}
\begin{equation}
\langle\sigma_{0}^{x}\sigma_{n}^{x}\rangle=
\begin{vmatrix}
G_{-1} & G_{-2} & \cdots & G_{-n}\\
G_{0} & G_{-1} & \cdots & G_{-n+1}\\
\vdots & \vdots & \ddots & \vdots\\
G_{n-2} & G_{n-3} & \cdots & G_{-1}%
\end{vmatrix}
\text{,}%
\end{equation}%
\begin{equation}
\langle\sigma_{0}^{y}\sigma_{n}^{y}\rangle=%
\begin{vmatrix}
G_{1} & G_{0} & \cdots & G_{-n+2}\\
G_{2} & G_{1} & \cdots & G_{-n+3}\\
\vdots & \vdots & \ddots & \vdots\\
G_{n} & G_{n-1} & \cdots & G_{1}%
\end{vmatrix}
\text{,}%
\end{equation}
and%
\begin{equation}
\langle\sigma_{0}^{z}\sigma_{n}^{z}\rangle=\langle\sigma^{z}\rangle^{2}%
-G_{n}G_{-n}\text{,}%
\end{equation}
where%
\begin{align}
G_{n}  =\int_{0}^{\pi}d\phi\frac{\tanh\left(\beta\omega_{\phi}\right)}{2\pi\omega_{\phi}}&\left\{ \cos\left(  n\phi\right)  \left(  1+\lambda
\cos\phi\right)  \right. \nonumber\\
&  \left.  -\gamma\lambda\sin\left(  n\phi\right)\sin\phi\right\}\text{.}%
\end{align}

%-------------------------------------------------------------------------------------------------------------------------------------

\bibliographystyle{elsarticle-num}
%\bibliography{}

\end{document}